\documentclass[twocolumn,showpacs,preprintnumbers,amsmath,amssymb,floatfix,prb,superscriptaddress]{revtex4-1}
\usepackage{gensymb,amsmath,amssymb}
\usepackage{graphicx}
\usepackage{dcolumn}
\usepackage{bm}
\usepackage{units}
\usepackage[usenames]{color}

\begin{document}


\title{Jahn-Teller distortions as a novel source of multiferroicity}

\author{Paolo Barone}
\affiliation{
Consiglio Nazionale delle Ricerche (CNR-SPIN), 67100 L'Aquila, Italy 
}

\author{Kunihiko Yamauchi}
\affiliation{
ISIR-SANKEN, Osaka University, 8-1 Mihogaoka, Ibaraki, Osaka, 567-0047, Japan
}

\author{Silvia Picozzi}%

\affiliation{
Consiglio Nazionale delle Ricerche (CNR-SPIN), 67100 L'Aquila, Italy
}


\date{\today}
\newcommand{\ba}{Ba$_{2}$CoGe$_{2}$O$_{7}$}
\newcommand{\bas}{BCGO}
\newcommand{\beq}{\begin{eqnarray}}
\newcommand{\eeq}{\end{eqnarray}}
\begin{abstract}
The Jahn-Teller effect is a fascinating and ubiquitous phenomenon in modern quantum physics and
chemistry.  We propose a
class of oxides  with melilite structure Ba$_2T$Ge$_2$O$_7$ ($T$=V,Ni)
where Jahn-Teller distortions are the main responsibles for the appearance of electric
polarization. At the hearth of the proposed mechanism lies the lack of inversion symmetry displayed by tetrahedrally coordinated transition-metal ions, which allows
for the condensation of polar Jahn-Teller distortions, at odds with octahedral coordination typical of conventional ferroelectric oxides with perovskite structure. 
Since the noncentrosymmetric local environment of transition-metal ions also activates the
proposed spin-dependent hybridization mechanism for magnetically-induced electric polarization, proper multiferroic phases with intrinsic magnetoelectric interaction could be realized in this class of low-symmetry materials. 
\end{abstract}

\maketitle


\section{Introduction}
Jahn-Teller (JT) distortions represent a universal mechanism by which spontaneous symmetry breaking may
occur in condensed-matter
systems. Colossal magnetoresistance in manganites has been explained invoking an essential role of the Jahn-Teller
effect\cite{cmm_book},
which has been also invoked in several high-temperature oxide and fullerene superconductors
\cite{jt_effect1,jt_effect2}. Even though JT distorsions are nonpolar in the perovskite $AB$O$_3$ structure displayed by many ferroelectric oxides, the ferroelectric transition in these systems has been also explained in terms of a pseudo Jahn-Teller (PJT) instability which can take place
whenever the vibronic coupling between the ground and excited states is sufficiently strong\cite{bersuker_book}.
The cooperative PJT effect in ferroelectrics has been known for a long time, nevertheless its possible realization in
multiferroic perovskite oxides, showing the simultaneous presence of magnetic and
ferroelectric ordering, has been only recently suggested\cite{bersuker_prl,ghosez,rondinelli,barone_dasgupta}.Interestingly, such mechanism
defies the empirical exclusion rule, according to which proper ferroelectricity and magnetism should be chemically incompatible and mutually exclusive\cite{spaldin}.

On the other hand, the quest for sizeable magnetoelectric (ME) effects has been pursued in the last decade mainly
among magnetically induced improper ferroelectrics, where the microscopic ME interaction is expected to be intrinsically large.
In this respect, a very interesting class of materials has been recently object of intense research activity, suggesting a
local origin of ME coupling in low-symmetry crystals through a  spin-dependent hybridization
mechanism\cite{arima.jpsj2007,jia.2007,xiang_mike.2011,tokura.prl2010,yamauchi.bcgo,yamauchi.cfo}. 
Essentially, the lack of inversion symmetry in the point group of the transition-metal ions may drive the appearance of local dipoles via an  anisotropic hybridization, modulated by the atomic spin-orbit coupling,  between the transition-metal ions and the surrounding oxygens. Materials belonging to the  melilite family, with
general formula $A_2 TB_2$O$_7$ ($A^{+2}$ alkaline metal, $T^{+2}$ transition metal, $B^{+2}=$Si, Ge), consisting
of $B_2$O$_7$ dimers linked by $T$O$_4$ tetrahedra, represent an
interesting playground, where the local properties of $T^{+2}$ cations with tetrahedral coordination could be explored in
details. Among these, several compounds 
have been already synthesized, i.e. Ba$_2$$T$Ge$_2$O$_7$ ($T=$Mn,Co,Cu)\cite{tokura.prl2010,murakawa.prb2012} and $A_2$CoSi$_2$O$_7$($A=$Sr,Ca)\cite{akaki_conf.2009,akaki.2012}. Due to the lack of inversion symmetry and to the quasi-two-dimensional character of their magnetic interactions, these materials have been predicted to host peculiar incommensurate magnetic spiral ordering\cite{zheludev.1996} and skyrmion excitations\cite{bogdanov.2002}, multiferroicity and highly non-linear magnetoelectric response\cite{yi.2008,tokura.prl2010,murakawa.prb2012,akaki.2012}, as well as magnetochiral\cite{bordacs.2013} and a giant directional dichroism in resonance with both electrically and magnetically active spin excitations\cite{kezsmarki.2011}. The latter phenomena have been mostly explained in terms of the local ME interaction discussed above; specifically, the multiferroic phase observed in Ba$_2$CoGe$_2$O$_7$ has been shown to have  an improper origin, the ferroelectric polarization being induced by the magnetic order.

In this respect, a yet unexplored possibility is  to consider JT instabilities and possible {\em proper ferroelectric transition} for $T$ ions showing degenerate electronic states in tetrahedral symmetry.
\color{black} 
The lack of inversion symmetry of the local tetrahedral environment allows for odd ionic displacements to couple with the degenerate electronic states and, therefore, for symmetry-allowed appearance of local dipole moments. 
\color{black} In principle, this could imply proper multiferroicity with
intrinsically large ME coupling \color{black} in the family of melilite oxides.
We notice that, despite the possible occurrence of polar JT distortions in crystals without local center of inversion has been known for a long time and experimentally revealed, e.g., in rare-earth oxides 
(such as  DyVO$_4$) or in praseodymium compouds (such as PrCl$_3$) \cite{kaplan_book, pelikh_1979, vekhter_1980, leask_1981, unoki_prl.1977,kishimoto_prb.2010, harrison_1976, taylor_1977}, so far its relevance in the field of multiferroics has been overlooked.
\color{black}
Based on these premises, we first discuss in Section \ref{sec2} the Jahn-Teller problem relevant for the tetrahedral units $T$O$_4$ appearing in the melilite crystal. On the basis of general symmetry considerations, we show
that polar JT distortions may locally develop in this structure when JT-active transition-metal ions are considered.
Density functional theory (DFT) calculations
are then used to propose novel melilite oxides where such polar JT effects can appear, possibly leading to ferroelectric, antiferroelectric or ferrielectric phases, as discussed in Sections \ref{sec3},\ref{sec4}. 

\section{Polar Jahn-Teller effect at tetrahedral sites of melilite structure}\label{sec2}
\begin{figure}[h!b]
\vspace{-0.4cm}
\begin{center}
\includegraphics[width=0.5\textwidth]{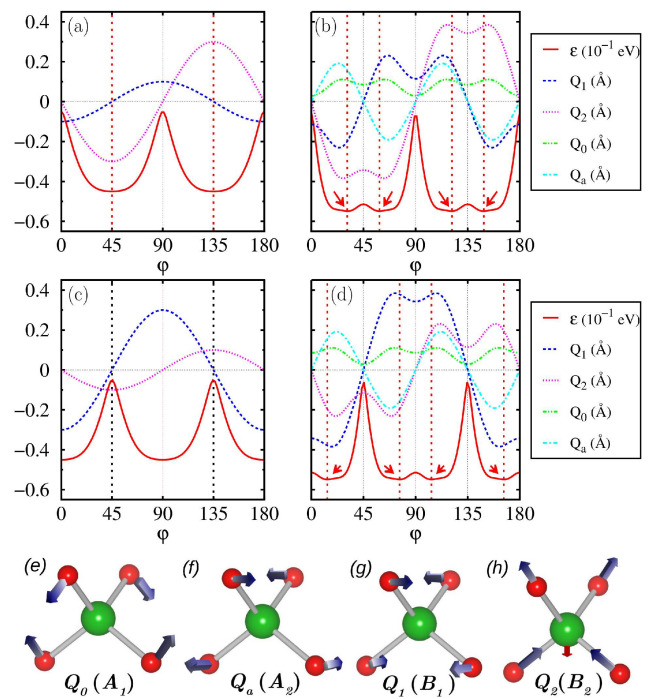}
\caption{(a)-(d): Typical evolution of JT modes --- shown in panels (e)-(h) --- and energy as a function of the electronic angle $\varphi$ for linear and quadratic couplings.
Parameters have been chosen as $K_1=K_2= K_a/2 =K_0/2 =1~ eV/$\AA$^2$,~ $F_1=0.3(0.1)~eV/$\AA~ , $F_2=0.1(0.3)~ eV/$\AA~ in top (bottom) panels, corresponding to cases $a$ and $b$ discussed in the main text. Quadratic coupling constants are set to zero in left panels (a),(c), while $L_1=L_2=0.9$ $eV/$\AA$^2$ and $G_1=G_2=0.5$ $eV/$\AA$^2$ in right panels (b), (d).
Arrows and vertical lines highlight
the energy minima. (c)-(f): Schematic representation of the symmetrized displacements in each tetrahedral unit, corresponding to totally symmetric displacements with symmetry $A_1$ (e) and $A_2$ (f), and to non-totally symmetric displacements with symmetry $B_1$ (g) and $B_2$ (h).} 
\label{fig1}
\end{center}
\end{figure}

Due to the quasi-layered structure of the melilite structure, each $T$O$_4$ appears to be slightly compressed along the $c$ axis, thus belonging to the tetragonal $D_{2d}$ group. As a consequence of the tetragonal crystal field, transition-metal $d$ states are split into nondegenerate $d_{z^2}, d_{x^2-y^2}$, two-fold $d_{yz}, d_{zx}$ and nondegenerate $d_{xy}$ states.
The
corresponding Jahn-Teller problem for degenerate $d_{yz}, d_{zx}$ is then expressed as $E\otimes(a_1+a_2+b_1+b_2)$.
By introducing the standard symmetrized displacements $Q_0(A_1),Q_a(A_2),Q_1(B_1)$ and
$Q_2(B_2)$, describing totally and non-totally symmetric displacements shown in Fig. \ref{fig1}(e)-(h), the vibronic matrix defined on the degenerate 
electronic states reads\cite{bersuker_book}:
\beq
W=\left(\begin{array}{cc}
\Delta_1 &\Delta_2\\
\Delta_2 &-\Delta_1
\end{array}\right),
\eeq
where 
\beq
\Delta_1&=&F_1 Q_1+L_2 Q_a Q_2 + G_1 Q_1 Q_0\nonumber\\
\Delta_2&=&F_2 Q_2 + G_2 Q_0 Q_2 + L_1 Q_1 Q_a,
\eeq
including both linear $F_1, F_2$ and quadratic
$G_i\equiv G(A_1\times B_i), L_i\equiv G(A_2\times B_i)$ couplings. Notice that the
non-totally symmetric
$Q_1$ and $Q_2$ modes describe a nonpolar and polar displacement, respectively, as highlighted in Figs. \ref{fig1}(g)-(h).
Upon inclusion
of the elastic term  $H_0=(1/2)\sum_i K_i Q_i^2$, the leading distortions as a function of all vibronic parameters
can be obtained by minimizing the corresponding energy $\mathcal{E}=
\frac{1}{2}\sum_i K_i Q_i^2\pm\sqrt{\Delta_1^2 +\Delta_2^2}$. Analytical solutions are found only when the quadratic
$L_i$ couplings are neglected; however, 
 all modes $Q_i$ of the full JT problem can be obtained as a function of the
stationary electronic states, namely of the mixing angle $\varphi$ through which the ground state is expressed as
$\vert \psi\rangle =\cos\varphi \vert yz\rangle + \sin\varphi \vert zx\rangle$\cite{opikpryce}.

When $L_i=0$, two kinds
 of energy minima are found, where only one of the non-totally symmetric modes with $B_1$ and $B_2$ symmetry is activated, either $Q_1=0, Q_2\neq0$ or $Q_1\neq0, Q_2=0$, depending on the ratio between linear vibronic and elastic energies $K_2F_1^2\gtrless K_1F_2^2$ as shown in Fig. \ref{fig1}(a), (c) and later on labeled as case $a$ and $b$, respectively \cite{bersuker_book}. In terms of the electronic wavefunctions, 
the two energy minima correspond to $\varphi_a=\pm\pi/4$ (case $a$) and $\varphi_b=0,\pi$ (case $b$), i.e. to a symmetric mixing of $d_{yz},d_{zx}$ states or to split levels with $d_{yz}$ or $d_{zx}$ unique character, respectively.
On the other hand, when $L_i\neq 0$ an effective coupling between nonpolar and polar modes appears through a term $(F_1L_2+F_2L_1)Q_aQ_1Q_2$  in the adiabatic potential energy $\mathcal{E}$. As a consequence, all distortion modes are found to be nonzero at energy minima, thus implying that a polar displacement is always symmetry-allowed, as shown in Fig. (\ref{fig1}); at the same time, JT split states display an asymmetric mixing of $d_{yz}$ and $d_{zx}$. Furthermore, the switching of $Q_2$ (hence of the local dipole) implies a change of sign of the product $Q_a Q_1$, consistently with the trilinear term coupling the two nonpolar and the polar modes.
As a consequence, the energy barrier related to the dipole switching can in principle be much reduced with respect
to the case of a single-mode instability [see, e.g., Fig. \ref{fig1} (d)]. 

\section{Jahn-Teller effect in {B\lowercase{a}$_2$N\lowercase{i}G\lowercase{e}$_2$O$_7$}}\label{sec3}
On the basis of the previous general analysis, we consider a hypotetical melilite oxide hosting JT-active $T$ ions, where the polar JT effect previously discussed is expected to appear. We performed DFT calculations resorting to the VASP code\cite{vasp} within the framework of generalized gradient approximation GGA-PBE, while the electronic correlation has been also taken into account by using the GGA+$U$ potential\cite{ldau} (with {\color{black} $U$=2, 4 or 6 eV} for $T$ ions). Electric polarization has been evaluated in the framework of the Berry-phase approach\cite{berry}. We started from the experimental
crystal structure of Ba$_2$CoGe$_2$O$_7$ belonging to $P\bar{4}2_1m$ space group\cite{crys.bcgo} and fully optimized it by substituting Co with the JT-active ion Ni$^{2+}$ while imposing different magnetic configurations of the transition-metal magnetic moments. We also considered V$^{2+}$, whose electronic configuration is physically equivalent to Ni$^{2+}$, but with occupied majority spins only. Indeed, results for both systems show qualitatively the same behavior, therefore we discuss here only the case of Ba$_2$NiGe$_2$O$_7$ (BNGO).
\begin{figure}[h!]
\vspace{-0.cm}
\begin{center}
\includegraphics[width=0.48\textwidth]{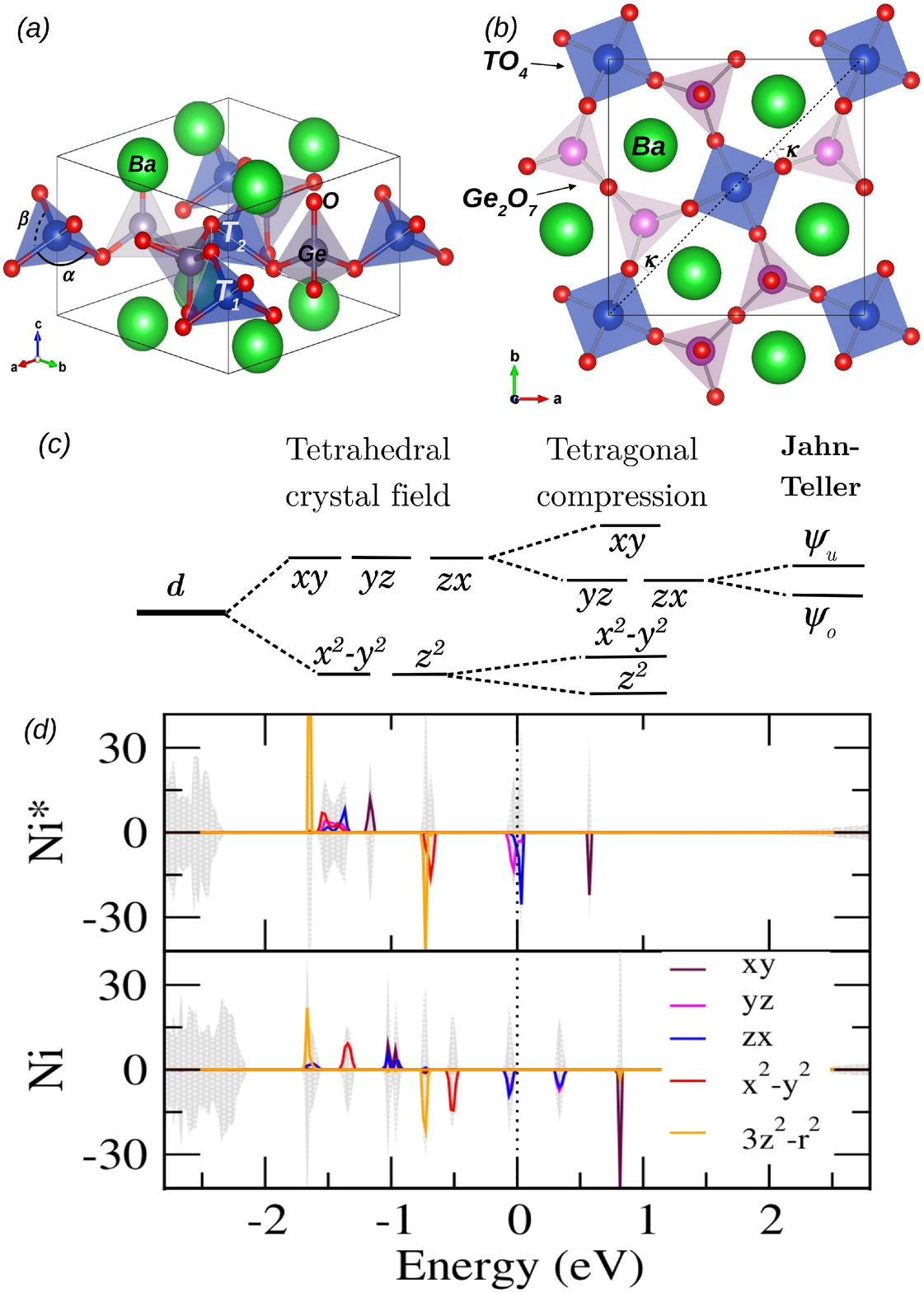}
\caption{(a) Melilite crystal structure of Ba$_2T$Ge$_2$O$_7$ with $P\bar{4}2_1m$ symmetry, showing layers of Ge$_2$O$_7$ dimers linked by $T$O$_4$ tetrahedra intercalated by Ba ions. The $T$O$_4$ tetrahedra are compressed along the $c$ axis, resulting in different O-T-O angles $\alpha, \beta$. (b) Top view of the crystal structure, highlighting two inequivalent $T$O$_4$ tetrahedra which are rotated about the $c$ axis of $\pm \kappa$, respectively. (c) Level structure of $T^{2+}$ metal ion in the melilite crystal. In the tetrahedral environment, $d$ orbitals split into lower $e_g$ and higher $t_{2g}$ manifolds; a tetragonal compression further split the $e_g, t_{2g}$ levels, leaving only twofold degenerated $d_{yz}, d_{zx}$ states. If these levels are half-occupied, as in the case of Ni$(d^8)$ or V$(d^3)$, a JT distortion can take place, removing the only left degeneracy. (d) Density of states (States/eV) decomposed into $d$ orbital
states and calculated within the bare GGA approach.
Ã
``Ni*''
refers to the parent-compound structure, belonging to space group $P\bar{4}2_1m$, 
while the optimized distorted structure with $Cmm2$ symmetry 
is labeled by ``Ni''.}\label{fig2}
\end{center}
\end{figure}

\begin{figure}[t]
\begin{center}
\includegraphics[width=0.48\textwidth]{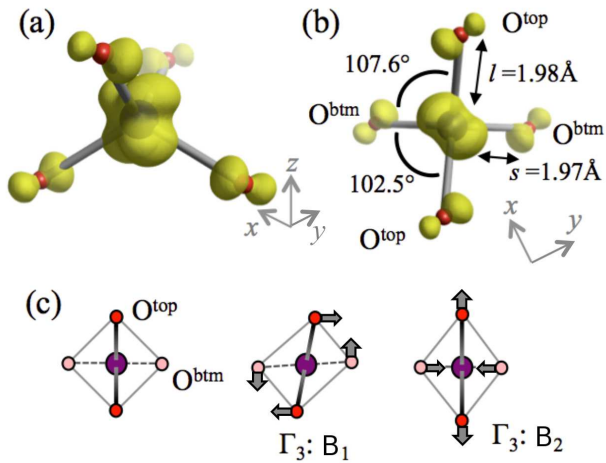}
\caption{Charge density plot (obtained within the GGA+$U$ approach, with $U$=4 eV) corresponding to the highest occupied
state of Ni-$d$ state hybridizing with tetrahedral O-$p$
states, (a) in perspective view in a local xyz frame, (b) projected
onto the xy plane. (c) Distortion modes of tetrahedral
O ions surrounding Ni ion, projected onto the xy plane, as obtained using the ISODISTORT program\cite{isodistort}.}\label{fig3}
\end{center}
\end{figure}

In Fig. \ref{fig2}(d), we show the density of states of BNGO  decomposed into $d$ orbital states, as obtained when a C-type antiferromagnetic configuration is imposed (similar results are obtained with other higher-energy magnetic configurations). When keeping the same space group $P\bar{4}2_1m$ of the parent compound, BNGO clearly shows a metallic behavior, arising from the twofold degeneracies of $(d_{yz}, d_{zx})$ orbital states which are half-filled in the minority spin manifold of Ni$^{2+}(d^8)$, being $e_g^{\uparrow\,2} t_{2g}^{\uparrow\,3} e_g^{\downarrow\,2} t_{2g}^{\downarrow\,1}$.  The degeneracy can then be lifted by a JT structural distortion; indeed, we found that
lowering the symmetry to $c$-centered $Cmm2$ space group leads to the opening of an energy gap in the BNGO compound, being $E_g = 0.3~ (2.3)~$eV at the bare GGA (GGA+$U$, $U$=4 eV) level, 
with a corresponding total energy gain of $78~(753)$ meV per formula unit (f.u.). 
{\color{black} We notice that the main effect of a finite $U$ in the GGA+$U$ approach is to induce a larger gap and a larger energy gain associated to the structural transition.}
\color{black} Both the Ni-O bond lengths and O-Ni-O bond angles are differentiated, as shown in Fig. \ref{fig3} (b).
Two short $s$ (long $l$) bonds can be identified, linking the Ni ion with upper (lower) lying oxygens in the tetrahedral cage, being $s$=1.95~\AA~ and $l$=2.00~\AA~ ($s$=1.97~\AA, $l$=1.98~\AA~ for $U$=4 eV). Furthermore, the O$^{top}$-Ni-O$^{btm}$ bond angles $\beta$, comprising upper- and lower-lying oxygens, differentiate in $\beta'$=$106.1^\circ\,$($107.6^\circ$ for $U$= 4 eV) and $\beta''$=$103.0^\circ\,$ ($102.5^\circ$ for $U$= 4 eV), suggesting that the JT-induced distortions do not consist only in a Ni offcentering. Indeed, 
the ionic displacement from the $P\bar{4}2_1m$ to the $Cmm2$ structure can be 
decomposed\cite{isodistort} into a totally symmetric $\Gamma_1$ mode, with total amplitude
$Q_{\Gamma_1}=0.02~$\AA, and a distortional (non-totally symmetric) mode, with total amplitude $Q_{\Gamma_3}=0.13~$\AA. The latter displacement mode can be further decomposed in two modes belonging to $B_1$ and $B_2$ symmetry representations, shown in Fig. \ref{fig3}(c) {\color{black} and corresponding to those shown in Fig. \ref{fig1}(g),(h), in perfect agreement with our previous qualitative analysis. These modes} display significant ionic displacements of O ions around the Ni cation, being $Q_{B_1}$=-0.10~\AA~ and $Q_{B_2}$=0.05~\AA. Upon structural distortion, the degenerate $yz,zx$ states are split into occupied  $0.61\vert yz\rangle - 0.79 \vert zx\rangle$ and unoccupied $0.79\vert yz\rangle +0.61 \vert zx\rangle$ states, implying a mixing angle $\varphi=37.7^\circ$. 
From the point of view of structural distortions, the $B_1$ mode, differentiating both the bond angle and the bond length, appears as the
largest JT mode; however, the electronic mixing angle appears closer to the value $\varphi_a$=$\pi/4$ of case $a$, suggesting that indeed the leading JT distortion arises from mode $B_2$, which is the main responsible for the $d$-orbital state splitting. {\color{black}On the other hand, the simultaneous activation of both $B_1$ and $B_2$ modes, alongside the nonzero amplitude of the totally symmetric displacement, points to a non-negligible quadratic couplings $L_i$, which is further confirmed by the asymmetric mixing of the $d_{yz}$ and $d_{zx}$ orbital states.}

\section{ Multiferroicity in Jahn-Teller melilites}\label{sec4}
If on the one hand the appearance of local dipoles in $T$O$_4$ cage can be understood in terms of JT effect, on the other hand the onset of an ordered ferro- (FE) or antiferroelectric (AFE) phase is related to non-local interactions between the different tetrahedral units. Since there are two $T$O$_4$ tetrahedra in the parent unit cell,
two possible structural configurations can be considered which realize the previously discussed local JT effect ; i) both $T$ ions offcenter downward or upward (FE, $P_c \gtrless 0$, space group $Cmm2$) , ii) $T$ ions offcenter in opposite directions
(AFE, $P_c = 0$, space group $P2_12_12_1$). As listed in Tab. \ref{Tab1}, the AFE structure in both compounds is energetically more stable by 20-30 meV/f.u..
The antiferro character of the cooperative JT interactions can be qualitatively understood in terms of the local coupling between on-site JT distortions. In fact, the sign of the polar distortion $Q_2$ is determined by the product $Q_a Q_1$, where the first nonpolar distortional mode is associated with a global rotation of the tetrahedral unit and the second to the antiphase rotation of O$^{top}$ and O$^{btm}$ around the tetrahedral $z$ axis, as shown in Figs \ref{fig1}(f) and (g), respectively. The sign of $Q_a$ is opposite in the two $T$O$_4$ units, which are rotated by an angle $\pm \kappa$ in the melilite crystal; on the other hand, $Q_1$ is expected to display the same sign, in order to minimize the energetic cost associated with distortions of the Ge$_2$O$_7$ dimers linking the JT-active units by inducing less asymmetric changes of the O-Ge-O bond angles. The opposite sign of $Q_a Q_1$ thus may explain the observed antiferro configurations of local dipoles associated with the $Q_2$ local modes. However, one can resort to a different JT ion occupying $T$ sites, such as V$^{2+}$. Indeed, when combining V and Ni ions, each carrying a different electric dipole, a ferrielectric configuration can be expected, as we show in Tab. \ref{Tab1}. On the other hand, the substantially local mechanism leading to the formation of electric dipoles suggests that ferroelectric behavior could be in principle attained when doping melilite oxides with JT-active ions; indeed, good-quality crystals of Ba$_2$Cu$_{1-x}$Ni$_x$Ge$_2$O$_7$ have been successfully synthesized up to Ni concentrations of $x=0.5$, their structural and magnetic characterization being under way\cite{Tatiana_published,Tatiana_private}.

\begin{table}[ht]
\vspace{-0.2cm}
\begin{tabular}{|lc|c|c|}
\hline
& &$\Delta E~$ (meV/f.u.) & $P_c$ ($\mu C/cm^2$)\\
\hline
BNGO&PE & 775.9 & --\\
&FE & 22.9 & 1.18 \\
&AFE & 0 & 0\\
\hline
BVGO& PE& 675.3 & --\\
&FE & 30.3 & 1.90 \\
&AFE &0&0\\
\hline
B(V,N)GO &PE & 339.7  & --\\
&strong FE & 189.7 & -10.85\\
&weak FE & 0 & 0.70\\
\hline
\end{tabular}\caption{Relative \color{black} GGA+$U$ ($U$=4 eV) \color{black} total energies and bulk polarization for \color{black} paraelectric metallic (PE), \color{black} FE and AFE configurations in BNGO, Ba$_2$VGe$_2$O$_7$ (BVGO) and an ordered half-doped Ba$_2$V$_{0.5}$Ni$_{0.5}$Ge$_2$O$_7$ (B(V,N)GO) hypotetical compound, where the parallel(antiparallel) configuration of inequivalent electric dipoles results in a strong(weak) FE phase. }\label{Tab1}
\end{table}

As for the magnetic properties, we found that both Ni and V oxides display an antiferromagnetic ground state, and therefore they can be considered as proper multiferroic materials. The in-plane exchange constant $J_{ab}$ appears to be the dominant antiferromagnetic interaction, while the out-of-plane exchange $J_c$ changes from ferro- to antiferromagnetic interaction when Ni is replaced by V, leading to C-type and G-type AFM configurations, respectively. In both cases, the estimated exchange anisotropy is rather strong, being $J_{c}/J_{ab}\lesssim 0.05$, thus displaying the quasi-2D character observed in known parent compounds. On the other hand, the most important contribution to the local spin-dependent hybridization mechanism mediating the reported large ME interaction in melilite oxides has been shown to arise from $d_{yz}, d_{zx}$ orbital states of the $T$ ions\cite{yamauchi.bcgo}. Since these states are partially filled in JT melilites, a significant spin-dependent modulation of charge density at lower- and upper-lying O ions through the asymmetric $pd$ hybridization is expected to further contribute to the local electronic polarization, i.e., to mediate a ME interaction. Due to the partial occupancy of the active $d$ states of $T$ ions, such ME interaction can be in principle of the same order of magnitude of that predicted for Co and Cu compounds, if not larger. Even though the relatively weak exchange interactions are responsible for very low N\'eel temperatures, signatures of such ME coupling are expected to be visible even in the paramagnetic phase, due to the localized nature of its microscopic origin.


\section{Conclusions}\label{sec5}

By combining symmetry-based JT analysis and DFT calculations, we put forward a novel mechanism for multiferroicity and theoretically predict the possibility of realizing proper FE and AFE phases in  melilite oxides. We have shown that JT instabilities of noncentrosymmetric units can lead to polar distortions and possibly mediate a (anti)ferroelectric transition of proper character \color{black} which could be signalled by large anomalies in the dielectric susceptibility. 
The cooperative JT origin of such structural transition is also known to lead to enhanced electrostrictive and magnetostrictive responses, that are expected to develop even if the predicted JT distortions display a dynamical, rather than static, character\cite{kaplan_book}.
\color{black} Remarkably, the JT polar distortions can be realized in the presence of a magnetic phase, thus circumventing the empirical exclusion rule between magnetism and ferroelectricity usually invoked for perovskite oxides. On the other hand, due to the proper nature of the predicted structural distortions, they are expected to take place at higher temperatures than those typically found for magnetically-induced improper ferroelectrics, at the same time displaying an intrinsic ME interaction of local origin that could be detectable even above the magnetic transition temperature. 

\acknowledgments

P.B. thanks Dr. R. Fittipaldi and Dr. A. Vecchione for useful and fruitful discussions. This work has been supported by CNR-SPIN Seed Project PAQSE002 and MIUR-PRIN Project ``Interfacce di ossidi: nuove propriet\`a emergenti,
multifunzionalit\`a e dispositivi per l'elettronica e l'energia'' (OXIDE). DFT calculations were performed using the facilities of the Supercomputer Center, Institute for Solid State Physics, University of Tokyo.

\end{document}